\documentstyle[12pt]{article}
\begin{document}
\def\theequation{\arabic{section}.\arabic{equation}}
\newcommand{\be}{\begin{equation}}
\newcommand{\ee}{\end{equation}}
\begin{titlepage}
\setcounter{page}{1}
\title{The $\omega \rightarrow \infty $ limit of Brans--Dicke theory}
\author{Valerio Faraoni \\ \\
{\small \it Inter--University Centre for Astronomy and Astrophysics 
(IUCAA)} \\ {\small \it Post Bag 4, Ganeshkhind, Pune 411 007, India}\\
{\small \it e--mail: faraoni@iucaa.ernet.in}
}
\date{}
\maketitle   
\begin{abstract}
The standard tenet that Brans--Dicke theory reduces to general relativity
in the 
$\omega \rightarrow  \infty $ limit has been shown to be false when the
trace of the 
matter energy--momentum tensor vanishes. The issue is clarified in a new
approach and the asymptotic behaviour of the 
Brans--Dicke scalar is rigorously derived.
\end{abstract} 
\vspace*{1truecm} 
\begin{center}  
To appear in {\em Physics Letters A}            
\end{center}     
\begin{center}  
\end{center}     
\end{titlepage}   \clearpage 
\setcounter{page}{2}
\section{Introduction} 

There is a surge of interest among theoretical physicists in Brans--Dicke
(BD) \cite{BD61} and scalar--tensor theories, motivated by the fact that
the association of scalar fields to the metric tensor seems unavoidable in
superstring theories \cite{GreenSchwarzWitten87}. In addition, the
scalar--tensor theories of which BD theory is the prototype exhibit a
conformal invariance property that mimics the conformal invariance of
string theories at high energies \cite{Cho92}--\cite{Brans97}, and is
applied below. Additional interest in BD and scalar--tensor theories comes
from the extended \cite{extended} and hyperextended \cite{hyperextended}
inflationary scenarios of the early universe. In spite of the fact that BD
theory is the oldest and best known alternative to general relativity
(GR), its essential features are not well understood. The standard tenet
(see e.g. \cite{Weinberg72}) that GR is obtained in the $\omega
\rightarrow \infty $ limit of BD theory has been shown to be false for
many exact solutions \cite{RomeroBarros92}--\cite{BanerjeeSen97}. In
addition, while it is believed that the BD field $\phi$ has the asymptotic
behaviour \be \label{1} \phi=\phi_0+ {\mbox O}\left( \frac{1}{\omega}
\right) \ee (where $\phi_0$ is a constant) as $\omega \rightarrow \infty$
\cite{Weinberg72}, for the above--mentioned solutions, one has instead the
asymptotic behaviour \cite{RomeroBarros92}--\cite{BanerjeeSen97} \be
\label{2} \phi=\phi_0+ {\mbox O}\left( \frac{1}{\sqrt{\omega}} \right) \;
.  \ee Only very recently it was realized that the anomaly in the $\omega
\rightarrow \infty$ limit is associated to the vanishing of the trace
$T={T^{\mu}}_{\mu}$ of the matter energy--momentum tensor $T_{\mu\nu}$
\cite{BanerjeeSen97}. This is a key point in the understanding of the
$\omega\rightarrow \infty$ limit;  the condition $T=0$ signals conformal
invariance, and it is natural to relate it to the conformal invariance
property of the gravitational part of the BD action.

When $T=0$, the entire BD action is invariant under a 1--parameter group
of conformal transformations ${\cal F}_{\alpha}$, and a change $\omega
\rightarrow \tilde{\omega}$ in the BD parameter is equivalent to a
transformation ${\cal F}_{\alpha}$ which moves a BD theory within an
equivalence class ${\cal E}$.  The $\omega \rightarrow \infty$ limit can
also be seen as a parameter change that moves BD theory within the same
class ${\cal E}$, and therefore it cannot reproduce GR, which does not
belong to ${\cal E}$. On the other hand, when $T \neq 0$, the conformal
invariance of BD theory is broken, the parameter change $\omega
\rightarrow \tilde{\omega}$ and the $\omega \rightarrow \infty$ limit
cannot be seen as a conformal transformation. One does not move within a
equivalence class which excludes GR, and the $\omega \rightarrow \infty$
limit can then reproduce GR.

Finally, the asymptotic behaviour (\ref{2}) of the BD scalar $\phi$ was
obtained as a order of magnitude estimate \cite{BanerjeeSen97}; using the
conformal transformation approach, Eq.~(\ref{2}) can be rigorously
derived.

Our notations and conventions are as follows: the metric signature is
--~+~+~+, the Riemann tensor is given in terms of the Christoffel symbols
by ${R_{\mu\nu\rho}}^{\sigma}= \Gamma^{\sigma}_{\mu\rho
,\nu}-\Gamma^{\sigma}_{\nu\rho ,\mu}+
\Gamma^{\alpha}_{\mu\rho}\Gamma^{\sigma}_{\alpha\nu}-
\Gamma^{\alpha}_{\nu\rho}\Gamma^{\sigma}_{\alpha\mu} $, the Ricci tensor
is $R_{\mu\rho}\equiv {R_{\mu\nu\rho}}^{\nu}$, and
$R=g^{\alpha\beta}R_{\alpha\beta}$. $\nabla_{\mu}$ is the covariant
derivative operator, $\Box \equiv g^{\mu\nu}\nabla_{\mu}\nabla_{\nu}$, and
we use units in which the speed of light and Newton constant assume the
value unity. 

\section{Conformal invariance}

We begin by considering the BD action in the Jordan conformal frame
\setcounter{equation}{0} \be S_{BD}=\frac{1}{16\pi}\int d^4x \sqrt{-g}
\left[ \phi R +\frac{\omega}{\phi} \, g^{\alpha\beta} \nabla_{\alpha}\phi
\nabla_{\beta}\phi \right] + S_{matter} \; , \ee where $S_{matter}$ is the
nongravitational part of the action, which is independent of $\phi$.  The
field equations are \be \label{4} R_{\mu\nu}-\frac{1}{2} g_{\mu\nu}
R=\frac{8\pi}{\phi} \, T_{\mu\nu} + \frac{\omega}{\phi^2} \left(
\nabla_{\mu}\phi \nabla_{\nu} \phi -\frac{1}{2} g_{\mu\nu}
\nabla^{\alpha}\phi \nabla_{\alpha}\phi \right) +\frac{1}{\phi} \left(
\nabla_{\mu}\nabla_{\nu} \phi-g_{\mu\nu} \Box \phi \right)  \; , \ee \be
\Box \phi =\frac{8\pi T}{3+2\omega} \; .  \ee Let us restrict, for the
moment, to consider the purely gravitational sector of the theory: under
the conformal transformation \be \label{6} g_{\mu\nu} \longrightarrow
\tilde{g}_{\mu\nu}=\Omega^2 g_{\mu\nu} \; , \ee where $\Omega (
x^{\alpha})$ is a nonvanishing smooth function, the Ricci curvature and
the Jacobian determinant $\sqrt{-g}$ transform as \cite{Wald84} 
\be
\label{7} \tilde{R}=\Omega^{-2} \left[ R+\frac{6\Box \Omega}{\Omega}
\right] \;\;\;\;\;\; , \;\;\;\;\;  \sqrt{-\tilde{g}}=\Omega^{4} \sqrt{-g}
\; , 
\ee 
and the BD Lagrangian density can be rewritten as follows 
\be
\label{9} {\cal L}_{BD} \sqrt{-g}= \sqrt{-\tilde{g}} \left[ \Omega^{-2}
\phi \tilde{R} -\frac{6\phi\Box \Omega}{\Omega^5} +\frac{\omega}{\Omega^2
\phi}\, \tilde{g}^{\mu\nu} \tilde{\nabla}_{\mu}\phi
\tilde{\nabla}_{\nu}\phi \right] \; .  
\ee 
By specifying the conformal
factor as 
\be \label{10} \Omega=\phi^{\alpha} 
\ee 
($\alpha \neq 1/2$) and
by redefining the scalar field according to 
\be \label{11} \phi
\longrightarrow \sigma= \phi^{1-2\alpha} \; , \ee one obtains \be
\label{12} {\cal L}_{BD} \sqrt{-g}=\sqrt{-\tilde{g}} \left[ \sigma
\tilde{R} +\frac{\tilde{\omega}}{\sigma} \, \tilde{g}^{\mu\nu}
\tilde{\nabla}_{\mu}\sigma \tilde{\nabla}_{\nu}\sigma \right] \; , \ee
where \be \label{13} \tilde{\omega}=\frac{\omega -6\alpha \left( \alpha -1
\right)}{\left( 1-2\alpha \right)^2} \; .  
\ee 
Hence, the gravitational
part of the BD action is invariant in form under the transformation given
by Eqs.~(\ref{6}), (\ref{10}), and (\ref{11}). These transformations
constitute a one--parameter Abelian group with a singularity in the
parameter dependence at $\alpha=1/2$. In fact, the consecutive action of
two maps ${\cal F}_{\alpha}$, ${\cal F}_{\beta}$ of the kind (\ref{6}),
(\ref{10}), (\ref{11}) with parameters $\alpha$ and $\beta$ gives a
transformation ${\cal F}_{\gamma}$ of the same kind with parameter $\gamma
\left( \alpha, \beta \right)= \alpha +\beta -2\alpha\beta$, and $\alpha,
\beta \neq 1/2$ implies $\gamma \neq 1/2$. The identity corresponds to the
transformation ${\cal F}_0 $ for $\alpha < 1/2 $. The inverse of the
transformation ${\cal F}_{\alpha}$ is the map ${\cal F}_{\bar{\alpha}}$
with $ \bar{\alpha}=-\alpha \left( 1-2\alpha \right)^{-1}$ for $\alpha <
1/2$. Finally, since $\gamma \left( \alpha, \beta \right)=\gamma \left(
\beta, \alpha \right)$, the group is commutative. 

If $M$ is a 4--dimensional smooth manifold, the BD spacetimes $\left( M,
g_{\mu\nu}^{( \omega )}, \phi^{( \omega )} \right)$ related by a
transformation
${\cal F}_{\alpha}$ constitute an equivalence class ${\cal E}$.

If one adds ordinary (i.e. other than the BD scalar) matter to the BD
action the conformal invariance is, in general, broken. However, under the
conditions $T_{\mu\nu}=T_{\nu\mu} $ and $T=0$ for the matter
stress--energy tensor, the conservation equation \be \label{14}
\nabla^{\nu} T_{\mu\nu}=0 \ee (which contains the dynamics of matter) is
conformally invariant \cite{Wald84}. Since $T_{\mu\nu} $ is not affected
by the field redefinition (\ref{11}), the total BD action is invariant
under the group of transformations (\ref{6}), (\ref{10}), (\ref{11}) if
$T=0$. In this case, a change of the BD parameter $\omega \rightarrow
\tilde{\omega}$ is equivalent to a transformation ${\cal F}_{\alpha}$ of
the kind (\ref{6}), (\ref{10}), (\ref{11}) for a suitable value of the
parameter $\alpha \neq 1/2$. Such a transformation ${\cal F}_{\alpha} $
maps the BD spacetime $\left( M, g_{\mu\nu}, \phi \right) $ corresponding
to the value $\omega $ of the parameter into another spacetime $\left( M,
\tilde{g}_{\mu\nu}, \sigma \right)$ corresponding to the value
$\tilde{\omega}$ of the BD parameter, which belongs to ${\cal E}$, and so
does the $\omega\rightarrow \infty$ limit. Hence, by performing this
limit, one cannot obtain GR solutions, because the latters do not belong
to the equivalence class ${\cal E}$ (GR is invariant under
diffeomorphisms, but not under conformal transformations). When matter
with $T\neq 0$ is added to the BD Lagrangian, the conformal equivalence is
broken, one no longer moves within the equivalence class ${\cal E}$ in the
$\omega\rightarrow \infty$ limit, and it is possible to obtain GR. 

Let us consider the singularity $\alpha =1/2$ in the function
$\tilde{\omega} \left( \alpha \right) $ given by Eq.~(\ref{13}); we
restrict the discussion to the range of values $\omega > -3/2$ (the case
$\omega <-3/2 $ is symmetric). $\tilde{\omega} ( \alpha )$ is singular at
$\alpha=1/2$, and has two branches; $\tilde{\omega}=\omega$ at $\alpha =0$
and $\alpha=1$, which correspond to the identity ${\cal F}_0$ in the group
of transformations (\ref{6}), (\ref{10}), (\ref{11}). The $\alpha
\rightarrow 1/2 $ limit corresponds to the $\tilde{\omega} \rightarrow
\infty$ limit of the BD parameter; when $\alpha=1/2 $, the conformal
transformation \be \label{100} g_{\mu\nu} \rightarrow
\tilde{g}_{\mu\nu}=\phi \, g_{\mu\nu} \ee and the scalar field
redefinition \be \label{101} \tilde{\phi}=\int \frac{\left( 3+2\omega
\right)^{1/2}}{\phi} \, d\phi \ee (instead of Eq.~(\ref{11}), which
becomes meaningless), recast the theory in the so--called Einstein
conformal frame (also called ``Pauli frame'' in
Refs.~\cite{Cho92,Cho94,Cho97}), in which the gravitational part of the
action becomes that of Einstein gravity plus a non self--interacting
scalar field as a material source \be \label{Einsteinaction} S=\int d^4x
\sqrt{-\tilde{g}} \left[ \frac{\tilde{R}}{16\pi} -\frac{1}{2} \,
\tilde{g}^{\mu\nu} \tilde{\nabla}_{\mu}\tilde{\phi}
\tilde{\nabla}_{\nu}\tilde{\phi} \right] \; .  \ee The transformation
(\ref{100}), (\ref{101}) has been known since the original BD paper
\cite{BD61} and it has later been generalized to scalar--tensor and
nonlinear gravity theories, and rediscovered a number of times (see
references in \cite{MagnanoSokolowski94}). The BD parameter disappears,
and the $\omega \rightarrow \infty $ limit cannot be considered: the
theory is already GR, apart from a violation of the equivalence principle
due to the anomalous coupling of the scalar to the energy--momentum tensor
of ordinary matter, if $T_{\mu\nu} \neq 0$ (\cite{MagnanoSokolowski94} and
references therein). 

Formally, BD theory with $\omega=-3/2$, which corresponds to the $\alpha
\rightarrow \pm \infty $ limit, is a fixed point of the transformation
(\ref{6}), (\ref{10}), (\ref{11}); in fact Eq.~(\ref{13}) gives
$\tilde{\omega}=\omega=-3/2$. However, the BD field equations are not
defined in this case.

\section{Asymptotic behaviour of the Brans--Dicke scalar as
$\omega\rightarrow \infty$}

The asymptotic behaviour (\ref{2}) of the BD scalar in the
$\omega\rightarrow \infty$ limit was derived in \cite{BanerjeeSen97} as a
order of magnitude estimate; using the conformal transformation approach,
it is straightforward to provide a rigorous mathematical derivation of
Eq.~(\ref{2}).

Under the condition $T=0$, any value of the BD parameter $\tilde{\omega}$
can be obtained starting from a fixed value $\omega$ (cf. Eq.~(\ref{13})).
Without loss of generality, we start from the value $\omega=0$ and solve
Eq.~(\ref{13}) with respect to $\alpha$ to obtain \setcounter{equation}{0}
\be \label{15} \alpha=\frac{1}{2} \left( 1 \pm
\frac{\sqrt{3}}{\sqrt{3+2\tilde{\omega}}} \right) \;  .  \ee When $
\tilde{\omega} \rightarrow \infty$, $\alpha \rightarrow 1/2$ and
Eq.~(\ref{11})  yields \be \label{16} \sigma \approx 1\mp \left(
\frac{3}{2\tilde{\omega}} \right)^{1/2} \, \ln \phi \ee as $
\tilde{\omega} \rightarrow \infty$. The ``old'' BD field $\phi$
corresponding to $\omega=0$ does not change in the limit, and the ``new''
BD field $\sigma$ has the asymptotic behaviour given by Eq.~(\ref{2}).

The asymptotic behaviour (\ref{16}) is the source of troubles in the
$\tilde{\omega} \rightarrow \infty $ limit of BD theory; since $
\nabla_{\mu} \sigma \approx \mp \left( 3/2\tilde{\omega} \right)^{1/2}
\nabla_{\mu} \ln \phi $, the second term in the right hand side of
Eq.~(\ref{4}) does not go to zero in the $\tilde{\omega} \rightarrow
\infty $ limit, and Eq.~(\ref{4}) does not reduce to the Einstein equation
with the same $T_{\mu\nu}$.

When $T \neq 0$, conformal invariance is broken, and the conformal
transformation approach cannot be applied. Instead, one has to resort to
the order of magnitude estimate of Ref.~\cite{Weinberg72} to derive
Eq.~(\ref{1}) instead of (\ref{2}). In the $ T \neq 0 $ case, we still
lack a rigorous mathematical derivation of Eq.~(\ref{1}). 

\section{Conclusions}

Only recently it was realized \cite{BanerjeeSen97} that the source of
troubles in obtaining GR as the $\omega \rightarrow \infty$ limit of BD
theory is related to the vanishing of the trace $T$ of the matter
energy--momentum tensor. The approach based on conformal transformations
allows one to understand the precise relation between the $\omega
\rightarrow \infty $ limit and the vanishing of $T$.  The failure to
obtain the correct GR limit when $T=0$ is explained in terms of the
invariance of the theory (when $T=0$) under the group of conformal
transformations ${\cal F}_{\alpha} $ given by Eqs.~(\ref{6}), (\ref{10})
and (\ref{11}). Since the $\omega \rightarrow \tilde{\omega} $ parameter
change (including the case $\tilde{\omega}=\infty$) simply moves BD theory
within the equivalence class ${\cal E}$, and GR does not belong to ${\cal
E}$, the attempts to obtain GR as the $\omega \rightarrow \infty $ limit
of BD theory are doomed to failure. It is only when matter with $T \neq 0$
is included into the BD action that this is possible, due to the breaking
of conformal invariance and to the fact that the change $\omega
\rightarrow \tilde{\omega}$ no longer moves the theory within a restricted
equivalence class.

The asymptotic behaviour of the BD scalar in the $\omega \rightarrow
\infty$ limit, under the condition $T=0$, is given by Eq.~(\ref{2}), which
receives a sound mathematical justification for the first time in the
conformal transformation approach. The latter is only applicable in vacuum
($T_{\mu\nu}=0$), or in the presence of matter satisfying the condition
$T=0$.

Finally, we point out a issue of potential interest: consider the
differential equation \setcounter{equation}{0} \be \label{L} L(a) f(
x^{\alpha} )=0 \; , \ee where $L(a)$ is a partial differential operator
depending on a parameter $a$. Let $L_0$ be the limit of $L(a)$ as $a
\rightarrow 0$, and let $f_0$ be the limit of a solution $f( x^{\alpha} )$
of Eq.~(\ref{L}) as $a \rightarrow 0$. If $\psi$ is a solution of the
equation $L_0 f=0 $, then in general one has $\psi \neq f_0$. Although the
$\omega \rightarrow \infty$ limit of the BD field equations reproduces the
Einstein equations when $T \neq 0$, it is not trivial that a BD exact
solution tends to the corresponding GR solution in the same limit. To the
best of our knowledge, this property of the BD field equations has not
been investigated in the literature, and also within the context of GR a
spacetime may not have a well--defined limit as some parameter varies
\cite{Geroch69}. This issue will be investigated in the future. 

\section*{Acknowledgment}

The author is grateful to L. Niwa for a careful reading of the manuscript.


\clearpage

\begin{thebibliography}{99}

\bibitem{BD61} C.H. Brans and R.H. Dicke, Phys. Rev. 124 (1961) 925.

\bibitem{GreenSchwarzWitten87} B. Green, J.M. Schwarz and E. Witten,
Superstring Theory (Cambridge University Press, Cambridge, 1987).

\bibitem{Cho92} Y.M. Cho, Phys. Rev. Lett. 68 (1992) 3133.

\bibitem{Cho94} Y.M. Cho, in Evolution of the Universe and its
Observational Quest, Proceedings, Yamada, Japan 1993, ed. H. Sato (Universal
Academy Press, Tokyo, 1994).

\bibitem{Turner93} M.S. Turner, in Recent Directions in Particle Theory
-- From Superstrings and Black Holes to the Standard Model, Proceedings of the
Theoretical Advanced Study Institute in Elementary Particle Physics, Boulder,
Colorado 1992, eds. J. Harvey and J. Polchinski (World Scientific, Singapore,
1993).

\bibitem{KolitchEardley95} S.J. Kolitch and D.M. Eardley, Ann. Phys. (NY) 241 (1995)
128.

\bibitem{Brans97} C.H. Brans, preprint gr--qc/9705069.

\bibitem{extended} D. La and P.J. Steinhardt, Phys. Rev. Lett. 62 (1989) 376; A.M.
Laycock and A.R. Liddle, Phys. Rev. D 49 (1994) 1827.

\bibitem{hyperextended} E.W. Kolb, D. Salopek and M.S. Turner, Phys. Rev. D
42 (1990) 3925; P.J. Steinhardt and F.S. Accetta, Phys. Rev. Lett. 64 (1990) 2740; 
A.R. Liddle and D. Wands, Phys. Rev. D 45 (1992) 2665; R. Crittenden and P.J.
Steinhardt, Phys. Lett. B 293 (1992) 32.

\bibitem{Weinberg72} S. Weinberg, Gravitation and Cosmology (Wiley, New York,
1972).

\bibitem{RomeroBarros92} C. Romero and A. Barros, Astrophys. Sp.
Sci. 192 (1992) 263.

\bibitem{RomeroBarros93a} C. Romero and A. Barros, Phys. Lett. A 173 (1993) 
243.

\bibitem{RomeroBarros93b} C. Romero and A. Barros, Gen. Rel. Grav. 25 (1993) 491.

\bibitem{PaivaRomero93} F.M. Paiva and C. Romero, Gen. Rel. Grav. 25 (1993) 1305.

\bibitem{Scheeletal95} M.A. Scheel, S.L. Shapiro and S.A. Teukolsky, Phys. Rev. D 51
(1995) 4236.

\bibitem{BanerjeeSen97} N. Banerjee and S. Sen, Phys. Rev. D 56 (1997) 1334.

\bibitem{Wald84} R.M. Wald, General Relativity (Chicago University Press, 
Chicago, 1984).

\bibitem{Cho97} Y.M. Cho, Class. Quant. Grav.  14 (1997) 2963.

\bibitem{MagnanoSokolowski94} G. Magnano and L.M. Sokolowski, Phys.
Rev. D 50 (1994) 5039.

\bibitem{Geroch69} R. Geroch, Comm. Math. Phys.  13 (1969) 180.

\end{thebibliography}
\end{document}